\documentclass[final,1p,times]{elsarticle} 
\usepackage{graphicx} 
\usepackage{amssymb} 
\usepackage{amsthm} 
\usepackage{lineno} 

\def\La{\langle}
\def\Ra{\rangle}
\newcommand{\eq}{{\,=\,}}

\journal{Nuclear Physics A} 
\begin{document} 

\begin{frontmatter} 


\title{Viscous hydrodynamics with bulk viscosity -- \\
uncertainties from relaxation time and initial conditions}

\author{Huichao Song and Ulrich Heinz}

\address{Department of Physics, The Ohio State University, Columbus,
         OH  43210, USA\\[-4ex]}

\begin{abstract} 
Bulk viscosity suppresses elliptic flow $v_2$, as does shear
viscosity. It can thus not be neglected when
extracting the shear viscosity from elliptic flow data. We
here explore uncertainties in the bulk viscous contribution
to viscous $v_2$ suppression that arise from presently
uncontrolled uncertainties in the initial value of the bulk
viscous pressure and its microscopic relaxation time.
\end{abstract} 

\end{frontmatter} 



\section{Introduction}
Recently, causal viscous hydrodynamics for relativistic heavy ion
collisions has been experiencing rapid development. Several groups
independently developed 2+1-dimensional viscous hydrodynamic codes
that implement longitudinal boost invariance but allow for arbitrary
dynamics in the two dimensional plane transverse to the beam
\cite{Romatschke:2007mq,Song:2007fn,Song:2007ux,Dusling:2007gi,%
Luzum:2008cw,Molnar:2008xj}. So far, most of these works have focussed
on studying the effects caused by shear viscosity and on developing
strategies for constraining the QGP shear viscosity to entropy density
ratio $\eta/s$ from experimental data \cite{Romatschke:2007mq,Song:2007fn,%
Song:2007ux,Dusling:2007gi,Luzum:2008cw,Molnar:2008xj,Song:2008si,%
Heinz:2008qm,Dusling:2008xj}. It was found that elliptic flow $v_2$ is
very sensitive to $\eta/s$ and that, due to the rapid expansion of the
fireballs created in heavy-ion collisions, even the minimal KSS bound
$\eta/s=1/4\pi$ \cite{Kovtun:2004de} leads to a strong suppression of
$v_2$ compared to the ideal fluid case \cite{Romatschke:2007mq,%
Song:2007fn,Song:2007ux,Dusling:2007gi,Luzum:2008cw,Molnar:2008xj}.
A first attempt by Luzum and Romatschke \cite{Luzum:2008cw} to extract
the QGP viscosity from experimental elliptic flow data, using viscous
hydrodynamics simulations, indicate that $\frac{\eta}{s}|_{QGP} < 5\times
\frac{1}{4\pi}$. To obtain a more precise value requires additional
theoretical effort in at least the following four directions (see
\cite{Song:2008hj} for references): \textbf{(1)} connecting viscous
hydrodynamics to a hadron cascade to properly account for effects from
the highly viscous hadronic stage; \textbf{(2)} including the effects from
bulk viscosity; \textbf{(3)} employing a more realistic equation of
state (EOS) that uses the latest lattice QCD data above $T_c$ matched to
a hadron resonance gas in partial chemical equilibrium below $T_c$, to
properly account for chemical freeze-out at $T_\mathrm{chem}\simeq 165-170
$\,MeV; and \textbf{(4)} a better treatment of the initial conditions that
not only aims to eliminate presently large uncertainties in the initial
fireball eccentricity but also properly accounts for pre-equilibrium
transverse flow and fluctuations in the initial fireball deformation and
orientation.

In this contribution we focus on point \textbf{(2)} and study effects from
bulk viscosity. A longer account of this work can be found in 
Ref.~\cite{Song:2009rh}. In \cite{Song:2008hj}, where we reported first results
and to which we refer the reader for additional details, we constructed and
used a function for the specific bulk viscosity $\zeta/s(T)$ that interpolates
between a "minimal" value well above $T_c$, based on strong-coupling
calculations using the AdS/CFT correspondence \cite{Buchel:2007mf}, to a zero
value in the hadron resonance gas well below $T_c$, using a Gaussian function
between these limits that peaks at $T_c$. (For a discussion of uncertainties
in $\zeta/s$ below $T_c$ see \cite{Kapusta:2008vb}.) We call this function
"minimal bulk viscosity". To study larger bulk viscosities, we made comparison
runs where this entire function was multiplied with a coefficient $C>1$. For
twice the minimal bulk viscosity we found that the viscous $v_2$ suppression
increases from 20\% (for a fluid that has only minimal shear viscosity
$\frac{\eta}{s}=\frac{1}{4\pi}$) to 30\% (for a fluid that additionally
features bulk viscosity at twice the "minimal" level, $C=2$). For a given
measured value of $v_2$, accounting for such a 50\% relative increase in the
viscous $v_2$ suppression translates into a reduction of the extracted shear
viscosity $\eta/s$ by roughly a factor $\frac{2}{3}$. Consequently, bulk
viscous effects cannot be ignored when extracting $\eta/s$ from experimental
data.

The analysis \cite{Song:2008hj} suggests that the main uncertainty
stems from insufficient knowledge of the peak value of $\zeta/s$ near the
phase transition\footnote{Currently, theoretical uncertainties for the peak
value of $\frac{\zeta}{s}$ near ${T_c}$ are very large. Extraction from
lattice QCD simulations gives a peak value around 0.7~\cite{Meyer:2007dy}.
This is more than 10 time larger than the string theory prediction based on
holographic models \cite{Gubser:2008yx}. A critical discussion of the lattice
QCD based extraction can be found in \cite{Moore:2008ws}.}. However, this is
only part of the story. Additional complications arise from the fact that the
expected peak in $\zeta/s$ near $T_c$ is due to rapidly growing correlation
lengths, associated with a "critical slowing down" of microscopic relaxation
processes near the phase transition. This probably leads near $T_c$ to a much
larger relaxation time $\tau_\Pi$ for the bulk viscous pressure than commonly
used for the shear viscous pressure (which near $T_c$ is of the order of
$0.5$\,fm/$c$, as obtained from kinetic theory \cite{Israel:1976tn,%
Baier:2006um,York:2008rr} or AdS/CFT~\cite{Bhattacharyya:2008jc}).
Large relaxation times can lead to strong memory effects, i.e. strong
sensitivity to the initial conditions for the bulk viscous pressure.
This is what we discuss here \cite{Song:2009rh}.

\section{Setup}
We use the code {\tt VISH2+1} \cite{Song:2007fn,Song:2007ux} to
solve (2+1)-dimensional transport equations for the energy momentum
tensor $T^{mn}$ and the bulk viscous pressure $\Pi$:
\begin{eqnarray}
\label{eq1}&& d_m T^{mn}=0\,, \qquad \qquad   T^{mn} = e u^m u^n -
(p+\Pi)\Delta^{mn}\,,
\\ \label{eq3}
        &&D \Pi
=-\frac{1}{\tau_{\Pi}}(\Pi+\zeta \partial{\cdot}u)
 -\frac{1}{2}\Pi\frac{\zeta T}{\tau_\Pi}
       d_k\left(\frac{\tau_\Pi}{\zeta T}u^k\right).
\end{eqnarray}
Here, $m$, $n$ denote components in $(\tau, x, y, \eta_s)$
coordinates, with covariant derivative $d_m$ (for details see
\cite{Heinz:2005bw}), $D\eq u^m d_m$ and
$\nabla^m\eq\Delta^{ml}d_{l}$ (where $\Delta^{mn} = g^{mn}{-}u^m
u^n$ is the projector transverse to the flow vector $u^m$) are the
time derivative and spatial gradient in the local comoving frame,
$\zeta$ is bulk viscosity, and $\tau_{\Pi}$ is the corresponding
relaxation time. For $\frac{\zeta}{s}(T)$ we use the
phenomenological construction described in \cite{Song:2008hj}. For
$\tau_\Pi$ we consider three choices: the constant values
$\tau_\Pi=0.5$ and 5\,fm/$c$, and the temperature dependent function
$\tau_\Pi(T)=\max[\tilde{\tau}\cdot\frac{\zeta}{s}(T),
0.1\,\mathrm{fm}/c]$ with $\tilde{\tau} = 120$\,fm/$c$. The last
choice implements phenomenologically the concept of critical slowing
down; it yields $\tau_\Pi\approx0.6$\,fm/$c$ at $T=350$\,MeV and
$\tau_\Pi\approx 5$\,fm/$c$ at $T_c$.

To study memory effects, we explore two different initializations for
the bulk viscous pressure: \textbf{(a)} Navier-Stokes (N-S) initialization,
$\Pi(\tau_0)=-\zeta\partial\cdot u$, and \textbf{(b)} zero initialization,
$\Pi(\tau_0)=0$. We use $\tau_0=0.6$\,fm/$c$. For all other inputs we
make standard choices as discussed in Refs.~\cite{Song:2007ux,Song:2008si}
and listed in the figure below.

\section{Bulk viscosity effects: uncertainties
from relaxation time and bulk pressure initialization}

\begin{figure}[t]
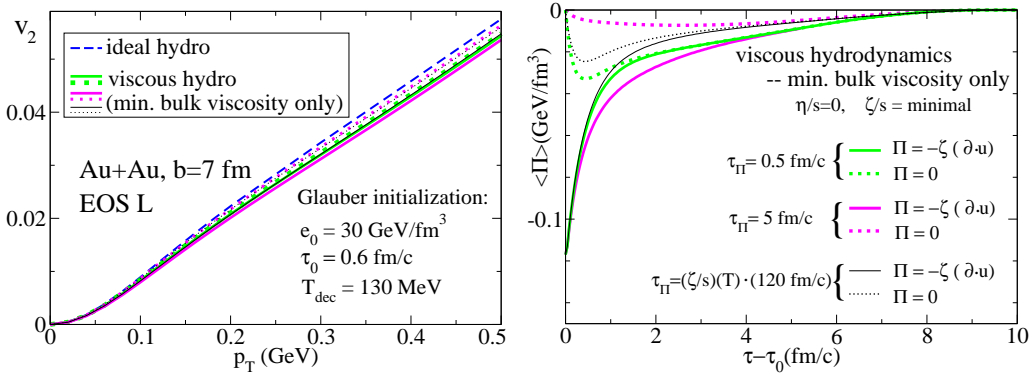

\vspace{-0.0cm}
\includegraphics[width=0.50\linewidth,clip=]{./Fig/v2_tauPi.eps} 
\includegraphics[width=0.49\linewidth,clip=]{./Fig/AvePi.eps}
\vspace{-0.6cm} 
\caption{\label{V2-Ave-PI}
(color online) {\sl Left:} Differential pion elliptic flow $v_2(p_T)$ from 
ideal and viscous hydrodynamics, including only bulk viscosity. 
{\sl Right:} Time evolution of the bulk pressure $\La\Pi\Ra$ averaged 
over the transverse plane (weighted by the energy density) from viscous 
hydrodynamics. Different curves correspond to different initializations 
and relaxation times, as indicated (see text for discussion).}
\vspace{-0.1cm} 
\end{figure}

The left panel of Fig.~\ref{V2-Ave-PI} shows the differential elliptic flow
$v_2(p_T)$ of directly emitted pions (without resonance decays) for
non-central Au+Au collisions at $b\eq7$\,fm, calculated from ideal
hydrodynamics and minimally bulk viscous hydrodynamics with identical
initial and final conditions. The different lines from viscous hydrodynamics
correspond to different relaxation times $\tau_\Pi$ and different
initializations $\Pi(\tau_0)$. One sees that these different inputs can lead
to large uncertainties for the bulk viscous $v_2$ suppression. For minimal
bulk viscosity, the $v_2$ suppression at $p_T=0.5$\,GeV ranges from
$\sim 2\%$ to $\sim 10\%$ compared to ideal hydrodynamics (blue dashed
line in the left panel).

For the shorter relaxation time, $\tau_\Pi= 0.5 \ \mathrm{fmc/c}$, the bulk
viscous $v_2$ suppression is insensitive to the initialization of $\Pi$, and
both N-S and zero initializations show $\sim 8\%$ $v_2$ suppression relative
to ideal fluids. The reason behind this becomes apparent in the right panel
showing the time evolution of the average bulk pressure $\La \Pi \Ra$. For
short relaxation times, $\La \Pi \Ra$ quickly loses all memory of its initial
value, relaxing in both cases to the same trajectory after about $1-2\,
\mathrm{fm/c}$ (i.e. after a few times $\tau_\Pi$). This is similar to what we
found for shear viscosity where the microscopic relaxation times are better
known and short ($\tau_\pi(T_c)\simeq 0.2-0.5\,\mathrm{fm/c}$) and where the
shear pressure tensor $\pi^{mn}$ therefore also loses memory of its
initialization after about 1\,fm/$c$ \cite{Song:2007ux}.

%

This changes if one accounts for the critical slowing down of the evolution
of $\Pi$ near $T_c$. If one simply multiplies the constant relaxation time
by a factor 10, setting $\tau_\Pi=5$\,fm/$c$, one obtains the dotted and
solid magenta lines in Fig.~\ref{V2-Ave-PI}. Now the bulk viscous $v_2$
suppression relative to the ideal fluid becomes very sensitive to the
initialization of the bulk viscous pressure: For zero initialization
$\Pi(\tau_0)=0$, the viscous $v_2$ suppression is very small (only $\sim 2\%$
at $p_T=0.5$\,GeV/$c$). The right panel shows that in this case the magnitude
of the (transversally averaged) bulk pressure evolves very slowly and always
stays small, leading to almost ideal fluid evolution. On the other hand,
if $\Pi$ is initialized with its Navier-Stokes value, which initially is
large due to the strong longitudinal expansion, it decays initially more
slowly than for the shorter relaxation time. Its braking effect on the
flow evolution is therefore bigger, resulting in much stronger suppression
of $v_2$ than for zero initialization, at $\sim 10\%$ slightly exceeding even
the viscous $v_2$ suppression seen for the tenfold shorter relaxation time.

The "critical slowing down" scenario with temperature-dependent $\tau_\Pi(T)$
(black lines) interpolates between the short and long relaxation times. As
for the fixed larger value $\tau_\Pi=5$\,fm/$c$, $v_2$ depends sensitively
on the initialization of $\Pi$, but for N-S initialization the viscous $v_2$
suppression is somewhat smaller than for both short and long fixed relaxation
times. The reasons for this are subtle since now, at early times, the bulk
viscous pressure $\Pi$ evolves on very different time scales in the dense core
and dilute edge regions of the fireball. As a result, for N-S initialization
the average value $\La\Pi\Ra$ is {\em smaller} in magnitude than for both short
and long fixed $\tau_\Pi$, throughout the fireball evolution (right panel, black
lines).

\section{Conclusions}
Relaxation times and initial values for the dissipative flows (bulk
and shear pressure) are required inputs in viscous hydrodynamic
calculations, in addition to the transport coefficients and the EOS.
Near $T_c$, the bulk viscosity $\zeta$ can exceed the shear viscosity
$\eta$ of the strongly interacting matter. If the relaxation time
$\tau_\Pi$ for the bulk viscous pressure $\Pi$ is short, it quickly
loses memory of its initial valaue, but the relatively large peak
value of $\zeta/s$ near $T_c$ can lead to a significant viscous
suppression of the elliptic flow $v_2$, competing with shear viscous
effects. If $\tau_\Pi$ grows rapidly near $T_c$, due to critical
slowing down, the bulk viscous suppression effects on $v_2$ depend
crucially on the initial value of $\Pi$: If $\Pi$ is zero initially,
bulk viscous effects on $v_2$ are almost negligible; if $\Pi$ is
initially large, however, as for the case of the N-S initialization,
it remains relatively large throughout the evolution, suppressing
the buildup of elliptic flow at a level that again competes with shear
viscous effects. Additional research on initial conditions and
relaxation times for the bulk viscous pressure is therefore necessary
for a quantitative extraction of $\eta/s$ from measured data.


\section*{Acknowledgments}
We thank K. Dusling and  P. Petreczky for enlightening discussions.
This work was supported by the U.S. Department of Energy under grant
DE-FG02-01ER41190.

\end{document}